\documentclass[referee]{aa}
\usepackage{graphics}
\begin{document}


\title{Asymptotic models of meridional flows in thin\\
    viscous accretion disks\thanks{Research supported by
    the Israel Science Foundation, the Helen \& Robert Asher
    Research Fund and by the Technion Fund for the Promotion of Research} }

\author{O. Regev\inst{1}
      \and L. Gitelman\inst{2}}

\offprints{Oded Regev,
\email{regev@physics.technion.ac.il}}

\institute{Department of Physics, Technion-Israel Institute of Technology,
  32000 Haifa, Israel
  \and Department of Mathematics, Technion-Israel Institute of Technology,
  32000 Haifa, Israel}

\date{Received: 9 July 2002 / Accepted: 25 September 2002}
\titlerunning{Meridional flows in accretion disks}

\abstract{
 We present the results of numerical integrations yielding
 the structure of and meridional flow in axisymmetric
thin viscous accretion disk models. The solutions are obtained by
simplifying and approximating first the equations, using
systematic asymptotic expansions in the small parameter
$\epsilon$, measuring the relative disk thickness. The vertical
structure is solved including radiative transfer in the diffusion
approximation. Carrying out the expansion to second order in
$\epsilon$ we obtain, for low enough values of the viscosity
parameter $\alpha$, solutions containing {\em backflows}. These
solutions are similar to the results first found by Urpin
(\cite{urpin}), who used approximations that are only valid for
large radii and the asymptotic analytical solutions of Klu\'zniak
\& Kita (\cite{kk}), valid only for polytropic disks. Our results
may be important for several outstanding issues in accretion disk
theory.

\keywords{accretion, accretion disks -- hydrodynamics -- stars:
formation -- stars: variables: general}
  }
\maketitle

\section{Introduction}
The Shakura-Sunyaev (hereafter SS) model of geometrically thin steady
accretion disks has extensively been used, since its publication
(Shakura \& Sunyaev \cite{ss}),
to model disks in a
variety of astronomical objects. In particular, accretion
disks, widely accepted to be present in mass transferring close binary systems
(like CV - cataclysmic variables) and in young, still accreting, stars (like classical T-Tauri),
were modeled using the SS paradigm and the calculated
emerging spectrum was compared with observations. The SS model, which is based
on the now famous $\alpha$ viscosity prescription, was found to be quite
successful in reproducing at least the gross features of the observations
and because of its relative simplicity and successful applicability it has achieved
the status of a text-book paradigm (see e.g. Frank, King \& Raine \cite{fkr}).

Two key ingredients facilitate the SS approach - the $\alpha$
viscosity prescription and the assumption that the vertical
extension of the disk is very small as compared to any significant
radial scale. The second of these is viable if the cooling of the
disk material (by radiative processes) is very efficient, so that
the sound speed $ c_s$ is very much smaller than the corresponding
swirling flow velocity $ r  \Omega$, which itself is close to the
Keplerian velocity. Choosing a representative typical value for
the radius $\tilde r$ and denoting the sound speed at that radius
by $\tilde c_s$, we obtain a natural small parameter for the
problem $\epsilon \equiv \tilde c_s/(\tilde \Omega \tilde r)$,
where $\tilde \Omega$ is the Keplerian angular velocity at the
representative point.

The SS model can naturally be understood in this context as the
lowest order (in $\epsilon$) approximation (see Regev 1983), but
it requires also some consideration of the vertical structure,
because the radiation (at least in the optically thick case)
escapes essentially in the $z$ direction. The common practice to
treat accretion disks using the SS paradigm has thus always been
accompanied by vertical "averaging" or "integration", that is, an
effective elimination of the $z$ dependence from the equations.
This gave a set of ordinary differential equations (ODE) in $r$
whose solutions provide the disk structure.

Studies wishing to model more accurately phenomena beyond the SS
model, like time-dependent behavior, radiatively driven mass loss,
various instabilities etc. have stretched the SS model to its
limits, but the trend was to stay within the vertical averaging
practice, because of the simplifications it introduces. For a
fairly recent review of such applications see e.g. Lin \&
Papaloizou (\cite{linpap}).

The question of the origin of the viscosity and the correct
implementation thereof has also been the subject of extensive
research efforts. The present day concensus identifies the
magneto-rotational instability (Balbus \& Hawley \cite{baha1}) as
the source of MHD turbulence, which ultimately produces angular
momentum transport (see Balbus \& Hawley \cite{baha2} for a
review). Although it seems that the classical $\alpha$
prescription is consistent with this process for vertically
averaged thin accretion disks (Balbus \& Papaloizou \cite{bapap}),
it is still unclear what the adequate model is to use in
multi-dimensional time-dependent studies, as it was recently
pointed out by Hawley, Balbus \& Stone (\cite{hbs}). This work
actually addresses nonradiative accretion flows (NRAF), that is,
flows in which the radiative cooling time is very long compared to
the time scales of interest. Previous studies in this context
introduced the so-called ADAF (Narayan \& Yi 1994, Abramowicz {\sl
et} al. 1995) and CDAF (see Abramowicz \& Igumenshchev 2001 and
references therein) giving rise to some controversy among the
different researchers.

Our work here deals with a thin, efficiently radiating, disk and
therefore seems to have no direct relation with the NRAF and
associated problems. Still the robustness of our results (which
hold in the polytropic and therefore adiabatic case as well) hints
that backflows may have some consequences also in NRAF.

In the study of thin disks it is  natural to employ systematic
asymptotic expansions in the small parameter $\epsilon$. Regev
(\cite{bl}), hereafter R83, introduced it for the purpose of
calculating the structure of an accretion disk-star boundary
layers and carried it out only in the lowest order. In that work,
matched asymptotic expansions were used, and it developed into a
full semi-analytic study of boundary layers in disks around
T-Tauri stars. However, not only were these expansions done only
up to the first order, but also the vertical structure of the disk
(and the boundary layer) was averaged out (using the usual
vertical integration procedures mentioned above).

Urpin (\cite{urpin}) was intetested in finding an approximation
for a steady accretion disk flow, beyond the SS solution, which
provides only the average inflow velocity. He was able to get the
two-dimensional approximate structure of the flow in the $r-z$
plane. In doing this he demonstrated that consideration of the
vertical gradients of the viscous stress tensor leads to a class
of solutions in which the flow direction in the midplane of the
disk may be opposite to that in the the layers around it, up to
the disk surface. In Urpin's work the viscous terms were not
treated consistently (neglected in one of the equations, but
retained in the other) and in addition a net zero angular momentum
transport was assumed. Therefore he obtained backflows for all
values of the viscosity parameter $\alpha$ and, in any case,  his
solutions are actually valid only for large radii.

Already the first attempts at a full two-dimensional
hydrodynamical numerical simulations of radiative viscous
accretion disks demonstrated that indeed the meridional flow
cannot often be well described by the vertically averaged value of
the velocity. In such calculations the result obviously depends on
the outer boundary conditions. Kley \& Lin (\cite{kleylin}) found
backflows near the midplane, similar to Urpin's solutions, but
have shown that if an inflow condition at the outer boundary is
assumed, a circulation pattern is obtained. More detailed studies
by R\'o\.{z}yczka {\em et} al. (\cite{rozyczka}) also exhibited
flows with regions of midplane backflows and with circulation
patterns.

Klu\'zniak \& Kita (\cite{kk}), hereafter KK, in a beautiful work,
carried out expansions in $\epsilon$ up to second order and
considered the full viscous flow dynamics, but neglected thermal
effects (which Urpin retained). Using a polytropic relation and an
inner zero torque boundary condition, they obtained global
analytical solutions with backflows. In these solutions the
location of the flow stagnation point on the equator $r_{\rm
stag}$ was found as a function of the viscosity parameter
$\alpha$.

In the present paper we would like to present the results of
asymptotic semi-analytical calculations of steady thin accretion
disks, similar to the ones obtained by KK, but with thermal
effects explicitly included via an energy equation, containing
viscous dissipation and radiative losses (treated in the diffusion
approximation). The inclusion of the energy equation will force us
to use some (quite straightforward) numerical ODE integrations and
thus the solutions will not have the closed analytical form of KK.

\section{Assumptions and equations}
We use the standard assumptions of steady state, axial and
equatorial symmetry, simplifying significantly the equations of
viscous fluid dynamics when they are expressed in cylindrical
coordinates.

Writing the equations in a non-dimensional form, that is, scaling
all the variables by their typical values (denoted by tilde),
brings out the non-dimensional constant $\epsilon$. This constant
was defined above and it reflects the disk thickness. It is thus a
natural expansion parameter for a perturbative treatment of
geometrically thin ($\epsilon\ll 1$) disks.

Denoting the radial, vertical and azimuthal velocity (in cylindrical
coordinates $r, z, \phi$) by $u$,$v$
and $r\Omega$ respectively, the statements of conservation of
mass, angular momentum, horizontal and vertical linear momentum
and energy, in steady state and with the assumption of axial symmetry, read:

\begin{equation}
\epsilon\,{1 \over r}\,\partial_r(r \rho u) +
\partial_z (\rho v) = 0,
\end{equation}
where $\rho$ is the mass density;

\begin{equation}
 {\epsilon\, u \over r^2}\,\partial_r (r^2 \Omega)
+ v\,\partial_z \Omega  =
 {\epsilon^2 \over  \rho r^3} \partial_r (\eta r^3\,\partial_r \Omega)
+  {1 \over \rho} \partial_z (\eta\,\partial_z \Omega),
\end{equation}
where $\eta \equiv \rho \nu$ is the viscosity coefficient. We note
in passim that this is, of course, the anomalous viscosity and it
is assumed that irrespective of its source it can be parametrized
using the usual (SS) $\alpha$ prescription, which allows us to
express the viscosity coefficient using the well known Ansatz for
the $r - \phi$ component of the viscous stress tensor

\begin{equation}
|\tau_{r, \phi}| = \alpha p,
\label{alphap}
\end{equation}

where $p$ is the pressure;
\begin{equation}
\epsilon^2\,u\,\partial_r u  +  \epsilon\,v\,\partial_z u  - \Omega^2 r = f^r_{\rm grav}
-\epsilon^2 {1 \over \rho}\,\partial_r p  + \epsilon\,f^r_{\rm vis},
\end{equation}

\begin{equation}
\epsilon\,u\,\partial_r v  + v\,\partial_z v =  f^z_{\rm grav} -
{1 \over \rho}\,\partial_z p + f^z_{\rm vis};
\end{equation}
and
\begin{equation}
{1 \over \gamma-1}(\epsilon\,u\,\partial_r T  +
 v\,\partial_z T) +
 T \left[\epsilon  {1 \over r}\,\partial_r
(r u) + \partial_z v  \right] =
 q^+_{\rm vis} - q^{-},
\end{equation}
where $T$ is the temperature and $\gamma$ the adiabatic constant
and where we have assumed that the disk material obeys the perfect
gas equation of state.

Some additional symbols appear in the above equations and they
are:
\begin{enumerate}
\item
The gravitational force (per unit mass) components, given by:
$$
f^r_{\rm grav}=- {1 \over r^2} \Bigl(1 +  {\epsilon^2 z^2 \over
r^2}\Bigr)^{-{3 \over 2}};\;\; f^z_{\rm grav}={ z \over r}\,
f^r_{\rm grav}.
$$
\item
The viscous force (per unit mass) components $f^r_{\rm
vis};~f^z_{\rm vis}$ and the viscous energy dissipation rate (per
unit mass) $q^+_{\rm vis}$. Since the expressions defining them
are somewhat lengthy we shall refrain from giving them explicitly here
and refer the reader to e.g. Tassoul (1978).
\item
The energy loss rate (per unit mass) $q^{-}$, whose form depends
on the physical conditions. Assuming only radiative losses we get
\begin{equation}
q^{-}=\Lambda {1 \over \rho} \left[{1 \over r}\,\partial_r\,(r
F^r)+\partial_z\,F^z \right],
\end{equation}
where $\Lambda$ is a non-dimensional constant and $F^r$, $F^z$ are
the appropriate radiative flux components. If the disk is
optically thick (as we shall assume in this work),the diffusion
approximation for radiative transfer is viable and the radiative
fluxes can be expressed in terms of the structure functions,
temperature gradient and the opacity in the usual way (see R83).
\end{enumerate}

\section{Lowest order solutions - the SS disk}

Expanding all the dependent variables in power series in
$\epsilon$ and denoting the succesive terms by the subscripts
$0,1,2\dots$ it is quite easy to see (R83, KK) that in the lowest
order we get
\begin{equation}
{1 \over r}\,\partial_r(r \rho_0 u_1) +
\partial_z (\rho_0 v_2) = 0,
\label{mass}
\end{equation}
because $u_0=v_0=v_1=0$,
\begin{equation}
r \rho_0 u_1 \partial_r(r^2 \Omega_0) = \partial_r(\eta r^3 \partial_r \Omega_0),
\label{am}
\end{equation}
\begin{equation}
\Omega_0=r^{3/2},
\label{rm}
\end{equation}
\begin{equation}
\partial_z p_0 = - \rho_0 z r^3,
\label{zm}
\end{equation}
\begin{equation}
\partial_z F_0 = \Lambda^{-1} \eta r^2 (\partial_r \Omega_0)^2,
\label{en}
\end{equation}
where $F_0\equiv F_0^z$.

The SS solution can be easily recovered from these equations (see
R83), but our aim here will be to obtain from them also the
vertical structure, needed for the next order approximation. Like
in the SS approach we vertically integrate equations (\ref{mass})
and (\ref{am}) and get
\begin{equation}
\int_{-\infty}^\infty r \rho_0 u_1 dz = const \equiv -\dot m,
\label{emdot}
\end{equation}
where the constant $\dot m$ is the nondimensional mass transfer rate, and
\begin{equation}
\dot m\, \partial_r(r^2 \Omega_0)=
-\partial_r(\mu r^3 \partial_r \Omega_0),
\label{zerotor}
\end{equation}
where $\mu \equiv \int_{-\infty}^\infty \eta dz$. Using now
(\ref{rm}) we can integrate equation (\ref{zerotor}) on $r$ and
use the condition that at some $r=r_+$ the zero torque condition
is satisfied, that is, $\Omega$ is at its extremum there (see KK).
The result is
\begin{equation}
\mu = {2 \over 3} \dot m \left(1 - \sqrt{r_+ \over r} \right).
\label{mu}
\end{equation}
The fact that a Keplerian angular momentum profile does not have
an extremum and that it has to be joined somehow to a
sub-Keplerian star gives rise to the boundary layer problem, which
we shall avoid here by looking only at radii sufficiently larger
than $r_+$.

To get the equations determining the vertical structure at each $r$ we use equations (\ref{zm})
and (\ref{en}) and substitute in the latter the Keplerian $\Omega_0$ (from equation \ref{rm}).
This gives
\begin{equation}
\partial_z p = -\rho z r^{-3},
\label{pver}
\end{equation}
\begin{equation}
\partial_z F = {9 \over 4 \Lambda} \eta r^{-3},
\label{fver}
\end{equation}
where we have dropped the subscripts $0$ from the functions, for convenience.
Treating the radiation in the diffusion approximation we get the additional equation
\begin{equation}
\partial_z T = -\kappa \rho T^{-3} F,
\label{tver}
\end{equation}
where $\kappa$ is the (properly non-dimensionalized) opacity, assumed known as a function of
$\rho$ and $T$.

This set of three equations, supplemented by the equation of state
$p =\rho T$, is sufficient in principle for the determination of
the vertical structure at each radial point $r$, but appropriate
boundary conditions are obviously also needed. This complicates
matters, because whereas the lower boundary can be simply set at
$z=0$, the upper one cannot be determined a priori. We chose to
employ the Eddington approximation for this boundary condition.
This approximation requires radiative equilibrium, that is no
energy sources in the outer layers, and to achieve it we tacitly
assume that the viscosity coefficient approaches zero for a small
enough optical depth. Actually, our form of the viscosity
parameter $\eta$ is obtained from the relation (\ref{alphap}),
which for the Keplerian (lowest order) $\Omega(r)$ gives
\begin{equation}
\eta={2 \over 3} \alpha r^{3/2} p.
\label{etafinal}
\end{equation}
Typically, $p$ decreases with $z$ and therefore so does $\eta$.
Thus using the Eddington approximation should be reasonable, at
least in the optically thick case, as long as we ignore possible
instabilities which may result in winds and coronae (see Shaviv,
Wickramasinghe \& Wehrse 1999).

 Explicitly, if we define the outer (vertical)
 limit of the disk and hence of our integration as $z_{\rm out}$ (see
 below), the integration of equation (\ref{fver}) from $z=-z_{\rm out}$ to
$z=z_{\rm out}$ gives
\begin{equation}
F(z_{\rm out})={9 \over 8 \Lambda} \mu r^{-3}={3 \over 4 \Lambda}
\dot m r^{-7/2}(r^{1/2}-r_+^{1/2}),
\label{fluxup}
\end{equation}
where the second equality follows from (\ref{mu}). This is the
upper boundary condition on the flux, but the value of $z_{\rm
out}$ has still to be specified. The lower boundary condition on
the flux is much simpler as it is just $F(z=0)=0$.

The Eddington approximation links the radiative flux and the
temperature at the base of the photosphere ($\tau_{\rm dim}=2/3$)
and this point is defined to be at $z=z_{\rm out}$, since the flux
above it is constant. In its non-dimensional form, using our
units, the Eddington condition reads
\begin{equation}
T^4(z_{\rm out})= {8 \over 3 \tilde \tau} F(z_{\rm out}),
\label{eddington}
\end{equation}
where $\tilde \tau=\tilde \kappa \tilde \rho \tilde h$ is our
optical depth unit.

The {\em three} equations (\ref{pver}-\ref{tver})  are thus
solved as a two-point boundary problem subject to the
{\em four} conditions
$F=0$ at $z=0$; (\ref{fluxup}), (\ref{eddington}) and $p=0$ at
$z=z_{\rm out}$, where the value of $z_{\rm out}$ itself
is also determined in an iterative way.

\begin{figure}
 \resizebox{\hsize}{!}{\includegraphics{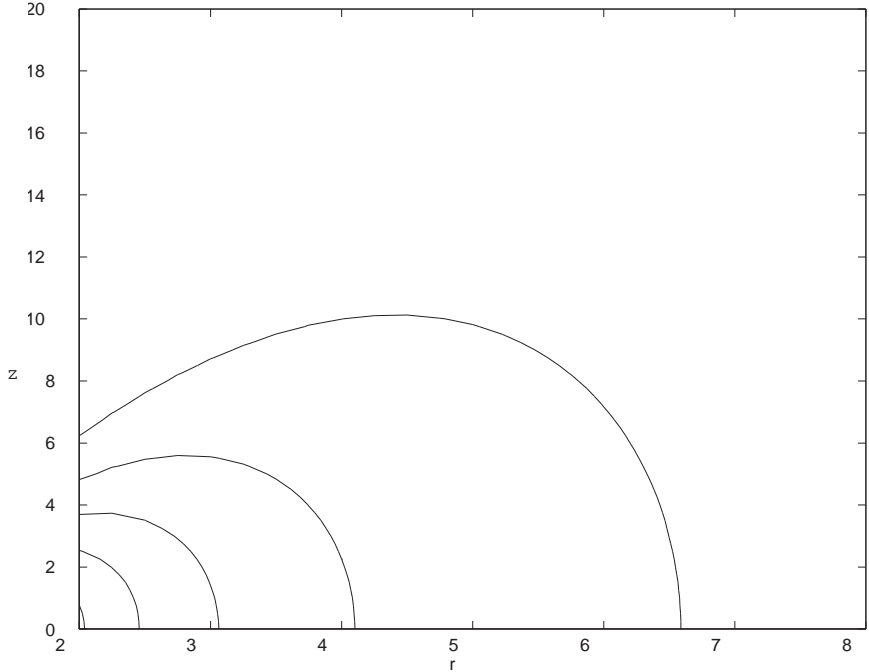}}
 \caption{Disk structure for a typical case.
 Contours of the density (in units of
  $\tilde \rho$) are plotted in the $r-z$ plane. Note that
  due to the different scalings the figure actually presents
  a vertically stretched disk by a factor of $\epsilon^{-1}\sim 50$}
 \label{contour}
\end{figure}

In figure \ref{contour} we show the result for $\rho(r, z)$ in a
typical case ($\alpha=0.2$ and electron scattering, i.e. constant,
opacity case). In this, and all subsequent figures the radial
coordinate is scaled by $\tilde r \equiv r_+$ and the vertical one
by $\tilde h \equiv \epsilon r_+$.

\begin{figure}
 \resizebox{\hsize}{!}{\includegraphics{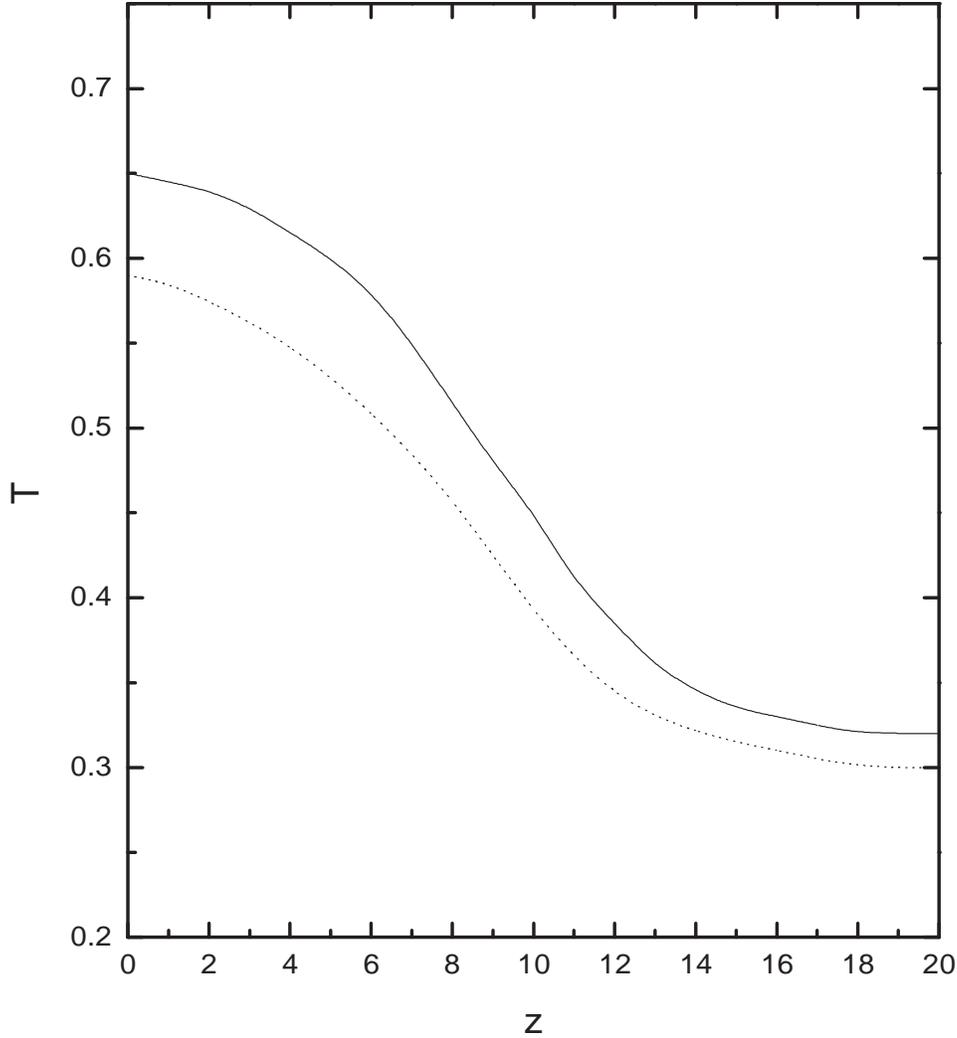}}
 \caption{Vertical temperature profiles for a typical radius
 in the disk.
 The non-dimensional temperature is shown as a function of
  $z$ (in units of $\epsilon r_+$) for $\alpha=0.4$ (solid line)
  and $\alpha=0.7$ (dotted line)}
 \label{temprofile}
\end{figure}

Figure {\ref{temprofile}} is the vertical temperature profile
$T(z)$ for a representative radius value and two values of
$\alpha$: $0.4$ (the solid line) and $0.7$ (the dotted line). As
we shall see in the next section, these two cases differ radically
in their meridional flow pattern, but we should stress that the
temperature profiles described here are calculated in the lowest
order and are, of course, not influenced by the flow.

The procedure outlined in this section provides the lowest order
values of the disk structure functions $p_0(r,z)$, $\rho_0(r,z)$
and so on, which will be needed to solve for the flow pattern in
the next order.

\section{Higher order solutions - the meridional flow pattern}
As explained above, the lowest order contributions to the
meridional velocity componenets are of order $\epsilon$ (i.e.
$u_1$) and $\epsilon^2$ (i.e. $v_2$) and it turns out that the
lowest order correction to the Keplerian angular velocity is of
order $\epsilon^2$ ($\Omega_2$) - see KK and Urpin (\cite{urpin}).
The relevant equations, following from the expansions and
originating from the mass, angular momentum and horizontal
momentum conservation equations are
\begin{equation}
\partial_r(r \rho_0 u_1) + r \partial_z(\rho_0 v_2)=0,
\label{mass2}
\end{equation}
\begin{equation}
r \rho_0 u_1 \partial_r(r^2 \Omega_0) = \partial_r(\eta r^3
\partial_r \Omega_0) + r^3 \partial_z (\eta \partial_z \Omega_2),
\label{am2}
\end{equation}
\begin{equation}
-2 r \rho_0 \Omega_0 \Omega_2 = {3 z^2 \over 2 r^4} \rho_0 -
\partial_r p_0 + \partial_z(\eta \partial_z u_1).
\label{rm2}
\end{equation}

Despite their appearance these equations can be viewed as ODE in
$z$ for any given $r$, since $p_0$, $\rho_0$ and $\Omega_0$, as
well as $\eta$ are already known functions, from the lowest order
solution for the vertical structure (as explained in the previous
section). In addition, equation (\ref{mass2}) can be decoupled
from the other two, because one of the variables, $v_2$, appears
only in it. Since equations (\ref{am2}-\ref{rm2}) are of second
order each, four boundary conditions are required. Two of them are
obvious and are the result of equatorial symmetry:
\begin{equation}
\partial_z u_1=0~~~{\rm and}~~~\partial_z \Omega_2=0;~~~{\rm at}~~~z=0,
\end{equation}
for any $r$. The other two conditions result from a regularity
requirement, that is, that $u_1$ and $\Omega_2$ remain bounded for
$z \to \infty$. In KK the density achieved a true zero for a
finite value of $z$, owing this property to the polytropic
assumption. In our case $\rho_0$ decays with $z$ and we have found
that physically meaningful results are to be expected only if we
require that both $\rho_0 u_1$ and $\rho_0 \Omega_2$ are
sufficiently small at $z_{\rm out}$. This expectation has been
strengthened by testing our algorithm on the polytropic case and
getting very good agreement with the analytical results of KK.

After the two equations (\ref{am2}-\ref{rm2}) are solved, $v_2$ follows from the solution
of the first order equation (\ref{mass2}) with the initial condition $v_2=0$ at $z=0$.

\begin{figure}
 \resizebox{\hsize}{!}{\includegraphics{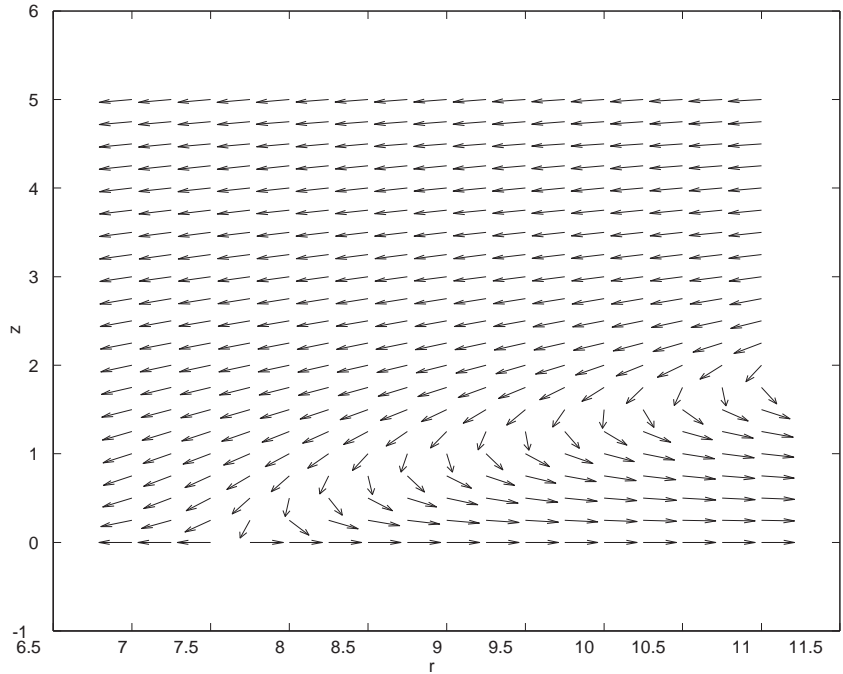}}
 \caption{Meridional flow pattern with backflow. The flow is represented by the
  direction of streamlines at a grid of points in the $r-z$ plane. Note that the scale
  of $z$ is "stretched" by a factor $\epsilon^{-1} \sim 50 $}
 \label{flow}
\end{figure}

The result of a representative calculation ($\alpha=0.4$ and
electron scattering opacity) is shown in Figure \ref{flow}. For
this case we obtain a meridional flow pattern with a backflow
around the equatorial plane. The stagnation radius is in this case
$r_{\rm stag}\sim 7.5 r_+$.

For other values of $\alpha$ the trend is similar to the one found
by KK, that is, $r_{\rm stag}$ growing with $\alpha$ and for
$\alpha>\alpha_{\rm crit}$ there are no solutions with backflows
($r_{\rm stag}=\infty$). While the disk structure depends somewhat
on the opacity law (we have experimented with electron scattering
opacity and Kramers opacity law), the {\em meridional flow
pattern} and the {\em stagnation radius} (see below) are quite
insensitive to it. In fact the relevant graphs (see below) for the
abovementioned two opacity laws look indistinguishable and
consequently we shall display only the electron scattering opacity
case.

\begin{figure}
\resizebox{\hsize}{!}{\includegraphics{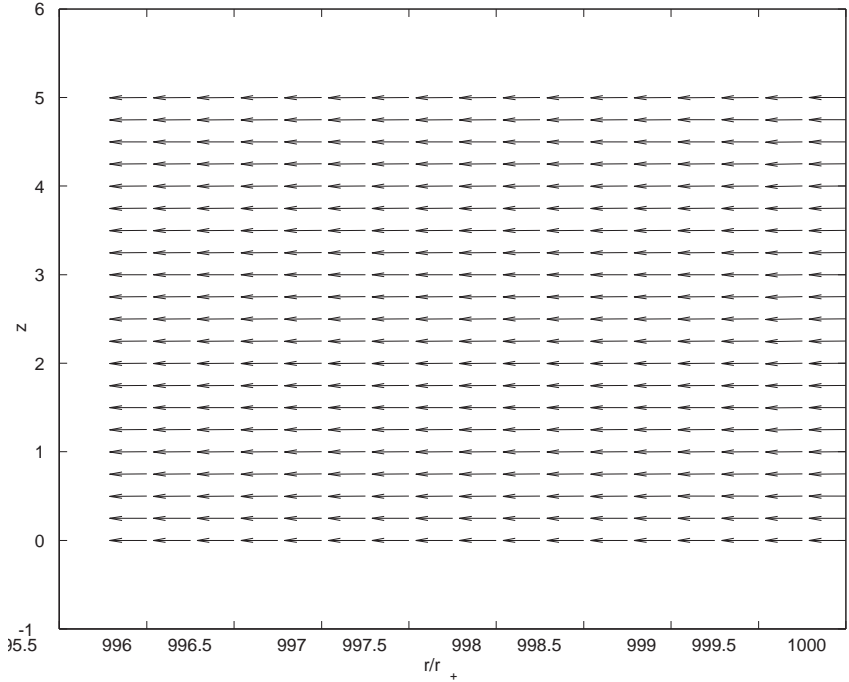}}
\caption{Meridional flow pattern with no backflow. The flow is
represented by the
  direction of streamlines at a grid of points in the $r-z$ plane. Note that the scale
  of $z$ is "stretched" by a factor $\epsilon^{-1} \sim 50 $}
 \label{flow2}
\end{figure}

A case in which the stagnation radius is so large that it is not
found up to $r \sim 1000 r_+$ is shown in fig. \ref{flow2}, where
$\alpha=0.7$. Note that even for these large values of $r$ the
flow is directed inward everywhere.

The determination of the value of $\alpha_{\rm crit}$ can be
achieved by calculating $r_{\rm stag}$ for a sequence of $\alpha$
values and finding where $r_{\rm stag}(\alpha)$ grows beyond any
reasonable bound. Reasonable, in this context, means a radius
beyond which the disk does not exist in a realistic astrophysical
setting. In any case, when $r$ is very large the present numerical
calculation cannot give reliable results because of the truncation
error.

\begin{figure}
 \resizebox{\hsize}{!}{\includegraphics{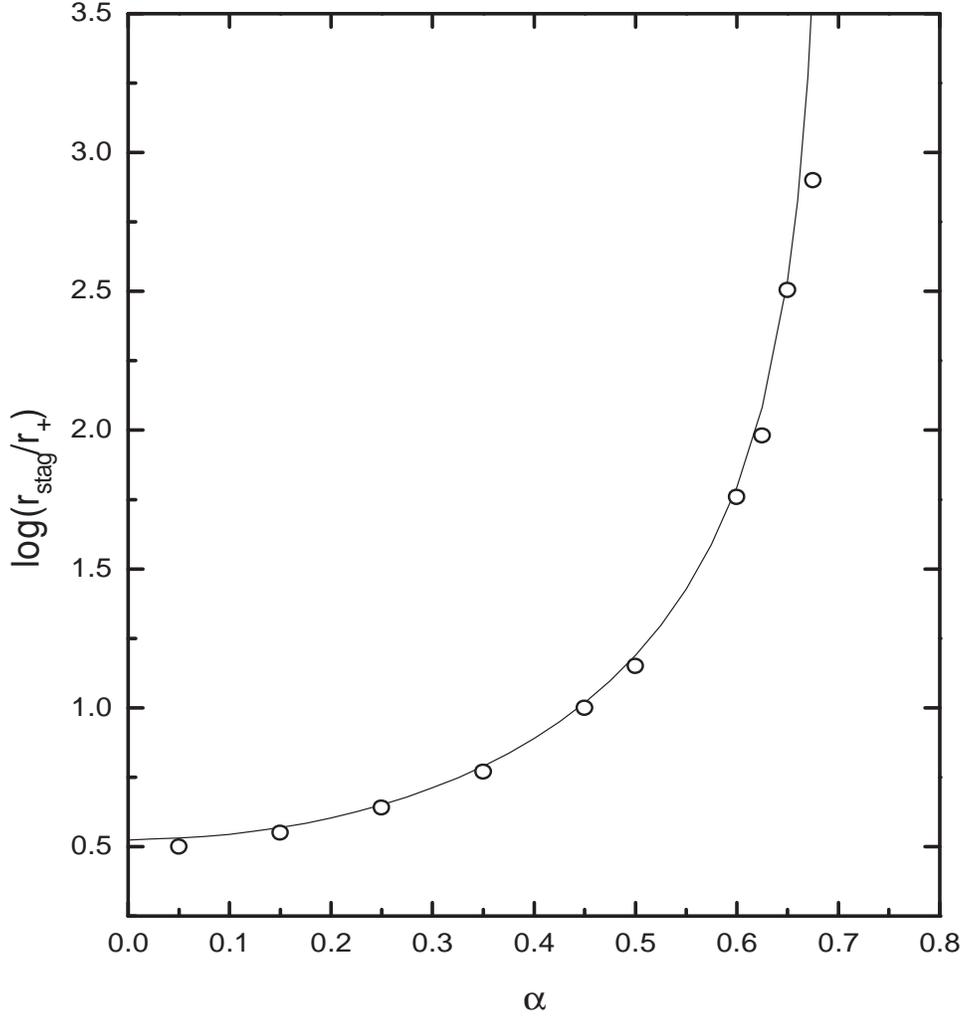}}
 \caption{Logarithm of the stagnation radius (in units of $r_+$) as a function of $\alpha$.
 The circles are our numerical results while the solid line is the KK analytical solution for a polytrope.}
 \label{stag}
\end{figure}

In Fig. \ref{stag} we show the stagnation radius for a number of
$\alpha$ values and clearly see the trend that it grows beyond any
reasonable bound for $\alpha \ga 0.7$. It is interesting to note
that our results are completely consistent with the analytical
result of KK (solid line) found for polytropes. The meridional
flow behavior seems to result thus from the dynamics rather than
from thermal processes in the disk.

\section{Summary}
In this work we have included thermal effects in the scheme,
previously proposed by Klu\'zniak and Kita (1997), to calculate
the vertical structure of thin accretion disks (including the flow
pattern) by perturbative analysis. The small parameter of the
expansions is $\epsilon$, reflecting the disk thickness. To the
lowest order in $\epsilon$ the SS solution is recovered and we
obtain also the vertical pressure and temperature structure. The
accretion flow is found from the next order and includes, for
small enough values of the viscosity parameter, equatorial
backflows.

These findings support the claim made by KK that the backflows are
of dynamical, rather than thermal, origin and probably result from
vertical gradients of the viscous stress tensor. Our result for
the flow is obtained numerically, using two explicit initial
condition at the disk mid-plane and two regularity conditions. The
latter complete the constraints needed for the solution of a
fourth order differential system. However, because these
constraints are independent of $r$, our solutions are obtained for
all $r$ values and cannot account for given detailed boundary
conditions on the flow at some $r=r_{out}$.

Still, it is conceivable that if such conditions are imposed in a
full numerical hydro simulation, our solution will be correct only
for an intermediate region, situated between $r_+$ and $r_{\rm
out}$, and not too close to these boundaries. Our solutions should
smoothly join the outer flow pattern, which has to be consistent
with the outer boundary conditions. For an outer boundary
condition of inflow along all $z$, this should give rise to
circulation patterns of the type observed in the numerical
simulations of radiative disks cited in the introduction.

Although our results are formally valid only for very small
$\epsilon$ and for radiative (as well as polytropic, see KK)
disks, it is reasonable (as is often the case in asymptotic
analyses of this type) that they carry over also to cases of
$\epsilon$ close to unity. This means that quasi spherical flows,
like the non-radiative ones (NRAF), may be endowed with
circulation patterns as well. Whether these results are directly
relevant to CDAF (if these flows are indeed proven to be viable)
remains unclear. It can, however, be definitely stated that ADAF
solutions using vertical averaging may not faithfully represent
radial energy transport when there are substantial backflows,
which naturally carry energy outwards in the hottest parts of the
disk. In addition, we think that an approach of the kind reported
here may provide useful tools for the understanding of some
complicated features of flow patterns and instabilities in
accretion disks.

\section*{Acknowledgement}
We would like to thank Wlodek Klu\'zniak for his remarks and
suggestions which helped to improve this paper.

%

\end{document}